# Chain connectivity and conformational variability of polymers: Clues to an adequate thermodynamic description of their solutions

## I: Dilute solutions

**Maria Bercea** [a)], **Maria Cazacu** [a)] and **Bernhard A. Wolf***

Institut für Physikalische Chemie und Materialwissenschaftliches Forschungszentrum
der Johannes Gutenberg-Universität Mainz, Jakob Welder-Weg 13, D-55099 Mainz, Germany

**Abstract**

This is the first of two parts investigating the Flory-Huggins interaction parameter $\chi$ as a function of composition and chain length. Part I encompasses experimental and theoretical work. The former comprises the synthesis of poly(dimethylsiloxane)s with different molar mass and the measurements of their second osmotic virial coefficients $A_2$ in solvents of diverse quality as a function of $M$ via light scattering and osmotic pressures. The theoretical analysis is performed by subdividing the dilution process into two clearly separable states. It yields the following expression for $\chi_0$, the $\chi$ value in range of pair interaction: $\chi_0 = \alpha - \zeta\,\lambda$. The parameter $\alpha$ measures the effect of contact formation between solvent molecules and polymer segments at fixed chain conformation, whereas the parameter $\zeta$ quantifies the contributions of the conformational changes taking place in response to dilution; $\zeta$ becomes zero for theta conditions. The influences of $M$ are exclusively contained in the parameter $\lambda$. The new relation is capable of describing hitherto incomprehensible experimental findings, like a diminution of $\chi_o$ with rising $M$. The evaluation of experimental information for different systems according to the established equation displays the existence of a linear interrelation between $\zeta$ and $\alpha$. Part II of this investigation presents the generalization of the present approach to solutions of arbitrary composition and discusses the physical meaning of the parameters in more detail.

* Corresponding author: Tel.: +49-6131-392-2491; fax: +49-6131-392-4640
  *E-mail address*: Bernhard.Wolf@Uni-Mainz.de
a) Permanent address: "Petru Poni" Institute of Macromolecular Chemistry, Iași, Romania

# List of symbols

| | |
|---|---|
| $A_2$ | second osmotic virial coefficient |
| $a$ | exponent of the Kuhn-Mark-Houwink equation |
| $c$ | concentration (mass/volume) |
| $G$ | Gibbs energy |
| $K$ | factor of the Kuhn-Mark-Houwink relation |
| $M$ | molar mass |
| $N$ | number of segments |
| $R$ | gas constant |
| $R_\theta$ | Rayleigh ratio at the scattering angle $\theta$ |
| $T$ | temperature |
| $V$ | volume |
| | |
| $\alpha$ | short for $\chi_{o,\text{f-c}}$ |
| $\beta$ | $= \chi_{\text{overlap. coil}}/\chi_{\text{isol. coil}}$ |
| $\zeta$ | conformational response to dilution $= \beta - 1$ |
| $\theta$ | measuring angle |
| $\Theta$ | theta temperature |
| $\kappa$ | $K_N \rho_2$ |
| $\lambda$ | $= \dfrac{1}{2} + \kappa N^{-(1-a)}$ |
| $\pi$ | osmotic pressure |
| $\rho$ | density |
| $\sigma$ | slope of $A_2$ vers $N^{-(1-a)}$ |
| $\varphi$ | volume fraction |
| $\Phi$ | average volume fraction of segments within the space defined by a polymer coil |
| $\chi$ | differential Flory-Huggins interaction parameter |
| $\chi_{\text{isol. coil}}$ | auxiliary Flory-Huggins interaction parameter describing the interaction within an isolated polymer coil |
| $\chi_{\text{overlap. coil}}$ | auxiliary Flory-Huggins interaction parameter describing the interaction within of two partly overlapping coils surrounded by pure solvent |
| $[\eta]$ | intrinsic viscosity |

*superscript*

| | |
|---|---|
| $-$ | molar quantity |
| $=$ | segment molar quantity |
| E | excess quantity |
| $\infty$ | infinitely long chains |

*subscripts*

| | |
|---|---|
| o | infinite dilution |
| 1 | solvent |
| 2 | polymer |
| c-r | conformational relaxation |
| f-c | fixed conformation |
| N | referring to polymer segments |
| w | weight average |
| $[\eta]$ | referring to information from intrinsic viscosities |

# I. INTRODUCTION

The majority of thermodynamic studies is still nowadays performed on the basis of the Flory-Huggins theory[1], despite its numerous and well known deficiencies, because of the straightforwardness of that approach. Some of the initial inadequacies could be eliminated easily, for instance by introducing composition dependent interaction parameters. Others, however, are so fundamental that no satisfying solution could be found so far. This is above all true for the collapse of the Flory-Huggins theory in the region of low polymer concentrations and for its disability to account for the experimentally observed[2] dependence of the Flory-Huggins interaction parameter, $\chi$, on chain length at high polymer concentrations. The present contribution (part I) addresses the first problem. The subsequent paper (part II) deals with the composition dependence of $\chi$. The two publications demonstrate how the just described inadequacies can be eliminated without loosing the simplicity of the theoretical equations.

The reason for the problems of the Flory-Huggins theory with highly dilute polymer solutions lies in the fact that the segments of a polymer chain cannot distribute among the entire volume of the system due their chemical bonds (chain connectivity). Numerous attempts have been made to solve this problem. Among them the excluded volume theory[3] is probably most sophisticated. However, even this approach is incapable of accounting for the indubitable experimental finding that the second osmotic virial coefficients, $A_2$, may *increase* (equivalent to a *decrease* of $\chi$) with rising molar mass of the polymer. The earliest report on this kind of uncommon behavior dates back half a century and is due to Flory and co-workers[4], who did, however, not evaluate the data accordingly so that they were not aware of that discrepancy.

In view of the principal difficulties of the Flory-Huggins theory with low polymer concentrations we abandon the normal ways of thinking. The strategy is no longer to start with the modeling of highly concentrated polymer solutions and to correct for the different situation prevailing in dilute solutions afterwards. Here we commence from the dilute side with two at least partly overlapping polymer coils, embedded in large surplus of pure solvent, and subsequently generalize the outcome to high polymer concentrations (part II).

Part I comprises an experimental and a theoretical part. The polymer we have chosen for the measurements is poly(dimethylsiloxane) (PDMS). There are two reasons for this selection: The possibility to check the validity of the above mentioned old literature data[4] and the fact that this polymer is liquid under ambient conditions and thus allows the determination of thermodynamic data in the entire range of composition. For theoretical reasons[5] the molecular weight of the PDMS samples should be as large and the chemical difference between end groups and monomeric units as small as possible. Hence we have synthesized the required samples ourselves. The theoretical section is based on an earlier approach[5], which is briefly recapitulated and slightly rephrased in view of its extension to high polymer concentrations in part II.

# II. EXPERIMENTAL PART

## A. Polymer synthesis and solvents

Tetramethyl-ammonium hydroxide pentahydrate, $(CH_3)_4NOH \cdot 5H_2O$ (Aldrich); octamethylcyclotetrasiloxane, $[(CH_3)_2SiO]_4$ (Petrarch Inc.) was dried over Na wire and then distilled under vacuum before use. Linear poly(dimethylsiloxane)s were synthesized therefrom by adapting a method described in the literature[6]. This consists in a bulk anionic ring opening polymerization of the octamethylcyclotetrasiloxane, using tetramethyl-ammonium hydroxide pentahydrate as catalyst[7] and dimethylformamide as promoter. The molecular weights of the obtained polymers were controlled by the amount of the catalyst. Dimethylformamide was added in order to increase the reaction rate by counter ion complexation and hindering the ion pairs formation[8]. First, the pre-established amount of tetramethyl-ammonium as a 10 wt% solution in water, was dehydrated by refluxing with benzene 15 min and vacuum distillation. Then octamethylcyclotetrasiloxane was added and the reaction mixture was stirred under inert atmosphere at 80 °C for 90 min. In the end, the temperature of the reaction mixture was elevated to 150 °C, where the catalyst is decomposed[6]. Unreacted or equilibrium cycles were removed by vacuum distillation at this temperature. The characteristics of the five samples used for the present study are given in Table 1 together with the measured second osmotic virial coefficients.

Table 1: Molar masses of PDMS and second osmotic virial coefficients in different solvents, as obtained from light scattering (LS) or osmosis (OS), plus refractive index increments

| Sample | $\overline{M}_w$ (kg/mol) | | $\overline{M}_n$ (kg/mol) | MEK 40°C | | TL 40°C | | n-Oct 40°C |
|---|---|---|---|---|---|---|---|---|
| | | | | $A_2 \cdot 10^4$ (mol·cm$^3$/g$^2$) | $10^2$ dn/dc (mL/g) | $A_2 \cdot 10^4$ (mol·cm$^3$/g$^2$) | $10^2$ dn/dc (mL/g) | $A_2 \cdot 10^4$ (mol·cm$^3$/g$^2$) |
| | (LS) | (GPC) | (GPC) | (LS) | | (LS) | | (OS) |
| 1 | 540 | 570 | 320 | 0.47$_6$ | -2.843 | 2.76 | 8.512 | 4.98 |
| 2 | 620 | 610 | 350 | 0.48$_2$ | -2.815 | 2.72 | 8.581 | 4.70 |
| 3 | 770 | 790 | 480 | 0.56$_6$ | -2.682 | 2.41 | 8.892 | 3.90 |
| 4 | 830 | 910 | 525 | 0.58$_6$ | -2.647 | 2.33 | 8.989 | 3.68 |
| 5 | 1090 | 1100 | 600 | 0.65$_5$ | -2.545 | 2.23 | 9.200 | 3.31 |

The solvents toluene (TL), methyl ethyl ketone (MEK), and n-octane (n-Oct) of pro analysis grade (purum ≥ 99 %) were purchased at Fluka and used without further purification.

**B. Methods**

Light scattering measurements were performed in TL and MEK at 40°C with a modified (SLS, G. Baur, Freiburg, Germany) static light scattering apparatus Fica 50 (Sofica, Paris) using a laser (632 nm) and measuring at angles from 20° to 145°. Polymer solutions in the range of 0.2 to 1 g/dL were prepared five days in advance and kept at 50°C in an oven. Prior to the measurements, they were filtered through a 0.45 μm membrane filter (Millipore) directly into the thoroughly cleaned optical cells (Hellma, Müllheim, Germany) and thermostatted in the light scattering apparatus for 15 min. The refractive index increments (dn/dc) measured at 40 °C are stated in Table 1. Because of very low refractive index increment of PDMS in n-octane (dn/dc < 0.01 mL/g), it was impossible to obtain reliable molar masses and second osmotic virial coefficients from light scattering experiments with this solvent. For that reason we have performed osmotic pressure measurements on the membrane osmometer OSMOMAT 090-B (Gonotec, Berlin, Germany), equipped with a cellulose triacetat membrane (10 000 Dalton). Linear plots were obtained for reduced osmotic pressure *vers.* concentration up to the highest concentration of 0.1 g/100mL.

GPC measurements were performed as described[9] earlier, using a Waters Chromatography system with two detectors (SC200 UV and RI-61) and Permagel $10^3 - 10^6$ polystyrene standard columns thermostatted at 25°C. Calibration was performed with narrow polydispersity polystyrene standards using the universal calibration for PDMS. The samples were eluted with toluene at a flow rate was 1 mL/min.

Intrinsic viscosities were determined at 40 °C by means of an Ubbelohde-type capillary viscometer (Schott-Geräte, capillary Ia). The solutions were freed of dust by filtering them through a commercial Millipore filter (0.45 µm). The flow times for the solvents at different temperatures were in all cases larger than 80 s. The reported intrinsic viscosities $[\eta]$ are the arithmetic average of the data resulting from the evaluation according to Huggins and to Kraemer.

### III. PRIMARY DATA

#### A. $M$ and $A_2$

Figure 1 gives an example for the light scattering measurements and Figure 2 for the osmotic experiments. The resulting information concerning molar masses and second osmotic virial coefficients is collected in Table 1. The information on the variation of $A_2$ with $M$ obtained in the described manner is also shown graphically in terms of the double logarithmic plots (Figure 3) usually used for such purposes[5].

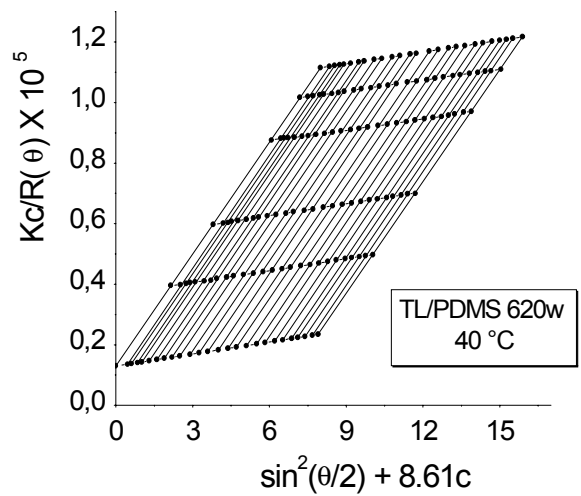

**Figure 1.** Zimm plot for the system TL/PDMS 620w at 40 °C; the numbers in the abbreviation of the polymer state the weight average molar mass in kg/mol. The measuring angles were varied in steps of 5° from 20 to 145°; the lowest concentrations was 0.250 and the highest 0.929 g/dL.

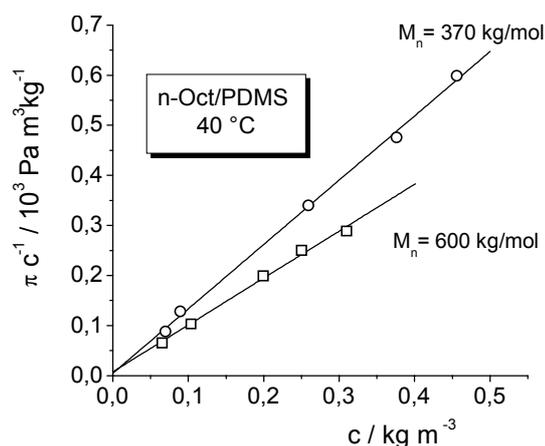

**Figure 2**. Reduced osmotic pressure $\pi/c$ as a function of polymer concentration for solutions of two PDMS samples in n-Oct.

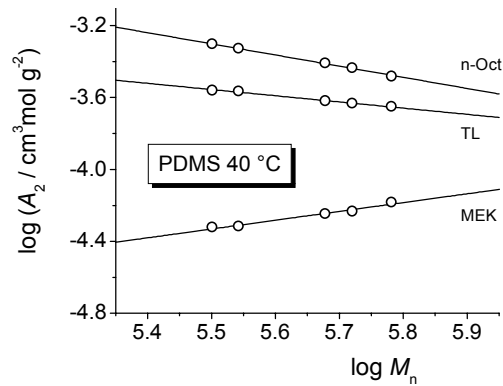

**Figure 3**. Conventional evaluation of the molecular weight dependence of the second osmotic virial coefficients $A_2$ at 40 °C in double logarithmic plots for the different solvents indicated in the graph.

## B. [$\eta$] and $k_H$

The results of capillary viscometry at 40 °C are presented in

Figure 4 for one polymer sample and the three solvents of different thermodynamic quality. Figure 5 shows the Kuhn-Mark-Houwink relations for all solvents and that for the system MEK/PDMS also at 20 °C; furthermore it includes the information for the same system but at 30 °C reported in the literature[4]. The resulting parameters are collected in Table 2 (p. 12).

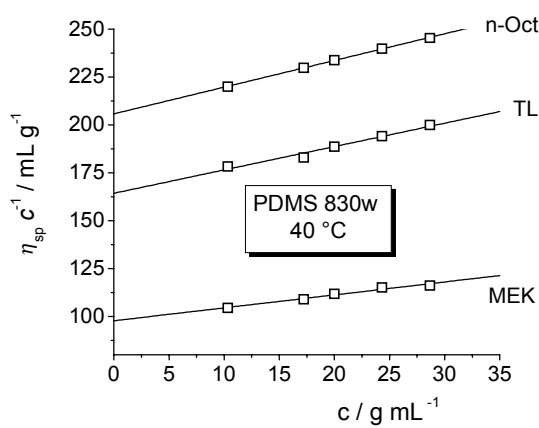

**Figure 4**. Huggins-plot for the determination of the intrinsic viscosities [$\eta$] of PDMS 830w in the different indicated solvents at 40 °C.

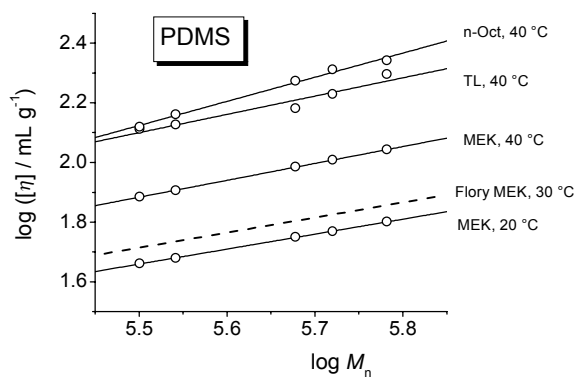

**Figure 5**. Kuhn-Mark-Houwink plot for PDMS. Solvents and temperatures are indicated in the graph; the dotted line represents literature data[4].

## IV. RESULTS AND DISCUSSION

The second osmotic virial coefficient $A_2$ yields the information required for the subsequent consideration. It is defined in terms of $\Delta \overline{G}_1^E$, the molar excess Gibbs energy of dilution, as

$$A_2 \equiv -\frac{\Delta \overline{G}_1^E}{RT\, c_2^2 \overline{V}_1} + \begin{pmatrix} \text{further terms of} \\ \text{series expansion} \end{pmatrix} \quad (1)$$

where $\overline{V}_1$ stands for the molar volume of the solvent and $c_2$ is the polymer concentration (mass/volume). According to the above formula $A_2$ measures the normalized effect of the addition of one mole of solvent to the dilute solution without changing its composition.

Data on the molecular weight dependence of $A_2$ are normally evaluated in double logarithmic plots as shown in Figure 3 for the present systems. Because of the failure of the excluded volume theory[3] and related approaches to account for the possibility of $dA_2/dM > 0$, we deal with the problem in a completely different manner[5], which considers chain connectivity and conformational variability of polymers[10].

The first feature has already been addressed in the introduction, the second consists in the ability of macromolecules of the present type to respond to changes in solvent quality by changing the population of the different conformers where the driving force is of course the minimization of the Gibbs energy of the entire system. The collective effect of this conformational variability can be easily observed in terms of an expansion or contraction of the polymer coils if the solvent quality changes with temperature. Analogous considerations also hold true if the solvent quality is kept constant and the polymer concentration varies. For instance, in thermodynamically good solvents the coils will expand upon dilution to realize more of the favorable contacts between polymer segments and solvent molecules. Newly added solvent is thus preferentially incorporated into the volume elements that already contain polymer segments instead of increasing number and/or size of the "lakes" of pure solvent. The subsequent considerations are based on the just described particularities.

First of all we translate $A_2$ (referring to interactions between polymer **molecules** and solvent molecules) into the Flory-Huggins interaction parameter $\chi$ (referring to interactions between polymer **segments** and solvent molecules). To this end we use the following phenomenological relation[11]

$$\chi_o = \frac{1}{2} - A_2 \rho_2^2 \overline{V_1} \qquad (2)$$

in which $\chi_o$ is the limiting value of $\chi$ for vanishing polymer concentration (composition range of pair interaction between the polymer *molecules*) and $\rho_2$ is the density of the polymer. We are now in a position to model the thermodynamic effects of the conformational variability by subdividing the dilution process, quantified by $A_2$, into two clearly separable steps[5] as briefly outlined in the next section.

**A. Dilution in two steps**

The spatial arrangement of the bonds connecting the segment of a polymer molecule will in the typical case not be identical for an isolated polymer coil and for a polymer molecule that overlaps at least in part with a second coil. In order to model this feature, we split the dilution process into two parts. The first step consists in the addition of solvent, retaining the conformations unchanged. This process unlocks exclusively contacts between *different* solute *molecules* (otherwise it would not contribute to $A_2$); this transaction does not change the number of contacts between segments belonging to the same molecule. To quantify the changes in the Gibbs energy of dilution associated with the opening of intermolecular contacts under the described boundary conditions, we introduce $\chi_{o,f-c}$, the Flory-Huggins interaction parameter for fixed chain conformation (previously called fixed dimensions), in the limit of high dilution. A second step is required if the equilibrium population of the different conformers (resulting in a particular spatial extension of the polymer coil) as it exists in isolated molecules differs from that realized in the presence of further polymer molecules. This term quantifies the relaxation of the system into its equilibrium state and is for instance indispensable for the many well established cases where the polymer coils expand upon dilution. The thermodynamic contribution of this conformational relaxation is called $\chi_{o,c-r}$. It results from a rearrangement of the local environment of the segments as consequence of the minimization of the Gibbs energy of the total system. The corresponding driving force will normally be of entropic nature (gain of entropy due to the additionally available space) but it might also enclose enthalpic contributions, which could for instance result from a better access of solvent molecules to favorably interacting sites of the monomeric unit after relaxation. According to the portrayed two-step mechanism we subdivide the experimentally accessible Flory-Huggins interaction parameter $\chi_o$ in two parts and write

$$\chi_o = \chi_{o,f-c} + \chi_{o,c-r} \qquad (3)$$

The meaning of the first summand appears rather clear-cut: In the absence of conformational changes resulting from dilution, $\chi_{o,f-c}$ constitutes the sole contribution to $\chi$ and can be directly measured. The situation is considerably more complicated with the second summand. In order to gain access to $\chi_{o,c-r}$ we perform an experiment in thought. We introduce an auxiliary parameter, namely $\chi_{isol.coil}$, quantifying the interaction between segments of an isolated polymer coil and the solvent molecules it contains. To allot a value to $\chi_{isol.coil}$ we insert a single, totally collapsed polymer molecule into a large surplus of good solvent. The macromolecule will take up solvent and swell until equilibrium is reached. Because of chain connectivity, which prevents the segments from leaving the coil, we can treat that process as the establishment of a "microphase equilibrium" between the solvent contained in the polymer coil and the surrounding pure solvent. In other words, the uptake of solvent will come to a halt as the chemical potential of the solvent inside the coil reaches that of the pure solvent existing outside. If we apply the Flory-Huggins equation to the two component system located inside an infinite surplus of pure solvent we can write[5]

$$\ln(1-\Phi_o) + \left(1 - \frac{1}{N}\right)\Phi_o + \chi_{\text{isol. coil}} \Phi_o^2 = 0 \quad (4)$$

where $\Phi_o$ represents the average volume fraction of the $N$ segments of an isolated macromolecule within the equilibrium "microphase", i.e. within the volume containing the polymer coil under equilibrium conditions. From the above relation it becomes immediately obvious that $\chi_{\text{isol. coil}}$ must assume large positive values to raise the chemical potential of the solvent inside the coil to that of the pure solvent. Furthermore it should be noted that increasing $\chi_{\text{isol. coil}}$ values correspond to decreasing $\Phi_o$, i.e. to higher coil expansion.

In view of the novelty of the approach formulated above, some additional thoughts, particularly concerning the meaning of the auxiliary parameter $\chi_{\text{isol. coil}}$, appear advisable. To this end we compare the *macroscopic* equilibrium established in an osmotic experiment with the *microscopic* equilibrium under consideration. In both cases two phases coexist, namely the pure solvent and a polymer solution. In the former case it is a semi-permeable membrane that impedes the solute *molecules* from spreading over the entire system. In the latter case it is chain connectivity, which constitutes the restriction hindering the *segments* of a polymer molecule to separate beyond a certain limit, which is determined by its chain length. For the macroscopic phase equilibrium the driving force for dilution (referring to solute *molecules*) is quantified by the osmotic pressure, $\pi$. This pressure is required to raise the chemical potential of the solvent in the solution to that of the pure solvent. For the microscopic phase equilibrium $\chi_{\text{isol. coil}}$ constitutes the analogous quantity, measuring the dilution tendency (per polymer *segment*) for an isolated polymer molecule. The auxiliary parameter $\chi_{\text{isol. coil}}$, like $\pi$, must become sufficiently large to increase the chemical potential of the solvent within the domain of the mixture to that of the surrounding pure solvent. The molecular reason for adverse interaction between polymer segments and solvent molecules within a given isolated polymer coil (large $\chi_{\text{isol. coil}}$) lies in unfavorable chain conformations of the polymer resulting from the opening of intersegmental contacts. Beyond a certain point of dilution (i.e. extension of the polymer coil) the loss in conformational entropy will raise $\chi_{\text{isol. coil}}$ to the value required for the establishment of equilibrium conditions. Given that chain connectivity constitutes the reason for the creation of microphases of the present kind, $\chi_{\text{isol. coil}}$ ought to depend on N. According to the current considerations $\chi_{\text{isol. coil}}$ should decline for otherwise constant thermodynamic conditions as N goes up (i.e. as the chain connectivity rises) and reach a finite limiting value for infinitely long chains.

Let us now compare $\chi_{\text{isol. coil}}$, referring to *intra*-molecular contacts between polymer segments, with the normal Flory-Huggins interaction parameter in the limit of high dilution, referring to *inter*-molecular contacts. Because of the fact that the purely enthalpic contributions should in good approximation be the same, irrespective of the type of contact that is opened upon dilution, these two parameters cannot be independent of each other. From the following considerations it will become obvious that a reduction in $\chi_o$ (improvement of solvent quality in terms of *inter*-molecular contacts) will – for a given chain length of the polymer - result in an augmentation of $\chi_{\text{isol. coil}}$. As the separation of segments by intruding solvent molecules becomes more favorable, the polymer chain can tolerate a larger loss in conformational entropy and expand more until equilibrium is reached. In other words: An improvement in solvent quality (reduction of $\chi_o$) increases the dilution tendency as measured by $\chi_{\text{isol. coil}}$.

The above considerations refer to the ultimate state of the dilution, i.e. to isolated coils. This situation cannot be studied experimentally by standard methods. The only directly accessible thermodynamic information refers to the separation of two polymer *molecules* and is quantified by the second osmotic virial coefficient. In order to extend our model to this more practical situation, we introduce $\chi_{\text{overlap. coil}}$, the interaction parameter within the "microphase" containing a pair of partly overlapping polymer coils. As already discussed, dilution will in the general case be associated with conformational changes so that $\chi_{\text{isol. coil}}$ and $\chi_{\text{overlap. coil}}$ need not be identical. In terms of these two parameters the contribution of conformational relaxation to the overall Flory-Huggins interaction parameter becomes

$$\chi_{o,c\text{-}r} = \chi_{isol.\,coil} - \chi_{overlap.\,coil} \qquad (5)$$

Because of the fact that $\chi_{\text{overlap. coil}}$ and $\chi_{\text{isol. coil}}$ cannot be independet of each other for a given system and because of the necessity that $\chi_{o,c\text{-}r}$ must become zero in the absence of conformational relaxation one can write

$$\chi_{overlap.\,coil} = \beta\,\chi_{isol.\,coil} \qquad (6)$$

so that we obtain the following expression for the contribution of conformational changes to the Flory-Huggins interaction parameter in the limit of high dilution

$$\chi_{o,c\text{-}r} = (1-\beta)\,\chi_{isol.\,coil} \qquad (7)$$

In order to give the factor $(1-\beta)$ a physical meaning and to simplify the nomenclature we introduce the conformational response $\zeta$ as

$$\zeta = -(1-\beta) \qquad (8)$$

The negative sign in front of the bracket was chosen to obtain positive parameter for the majority of systems for which $(1-\beta)$ results negative. The conformational response $\zeta$ becomes zero for theta systems, in agreement with the observation that conformational changes upon dilution are absent, as testified on a larger scale by the fact that the coil dimensions do not depend on composition.

In order to quantify the term $\chi_{o,c\text{-}r}$ on the basis of Eq. (7) we calculate $\chi_{\text{isol. coil}}$ according to Eq. (4). To this end $\Phi_o$, the average volume fraction of polymer segments within an isolated coil, must be known. Because of the thermodynamic nature of $\Phi_o$ one should use adequately determined coil dimensions, for instance that obtained from light scattering measurements. However, such data are scarce as compared with the abundant information on intrinsic viscosities $[\eta]$. Intrinsic viscosities measure the specific hydrodynamic volumes of isolated coils. If quoted as mL/g, $[\eta]$ indicates the volume of isolated coils (mL) formed by 1 g of the polymer. The molar volume of isolated coils is therefore given as $[\eta]\,M$. By means of the expression $N\bar{V}_1$ for the molar volume of the polymer itself (the solvent molecule defines the volume of a segment) we obtain the following equation for the volume fraction of segments contained in an isolated coil

$$\Phi_o = \frac{1}{[\eta]\rho_2} \qquad (9)$$

Insertion of Eq. (9) into Eq. (4) and rearrangement yields

$$\chi_{[\eta]} = \frac{1}{2} + \frac{[\eta]\rho_2}{N} \qquad (10)$$

for $\chi_{[\eta]}$, the interaction parameter between solvent molecules and polymer segments inside an isolated coil (the subscript of $\chi$ specifies the source of information).

The intrinsic viscosity, required for the calculation of $\chi_{[\eta]}$, can be obtained for the $N$ values of interest, by means of the Kuhn-Mark-Houwink relation

$$[\eta] = K_N N^a \quad (11)$$

where the parameter $K_N$ is related to the tabulated parameters $K$ referring to the molar mass of the polymer, instead of the number of segments, by the expression

$$K_N = K\left(\frac{\rho_2}{\rho_1} M_1\right)^a \quad (12)$$

in which $\rho$ stands for the densities of the components and $M_1$ for the molar mass of the solvent. Inserting the above Eqs. into Eq. (7) yields

$$\chi_{o,c-r} = -\zeta\left(\frac{1}{2} + \kappa N^{-(1-a)}\right) \quad (13)$$
$$\text{with} \quad \kappa = K_N \rho_2$$

For a given conformational response $\zeta$, the contribution of the conformational relaxation to $\chi_o$ does still depend on chain length. The effects (absolute values) are largest for the shortest chains and reach a limiting as $N \to \infty$. How this boundary value is approached depends on the thermodynamic quality of the solvent via the Kuhn-Mark-Houwink parameters $\kappa$ and $a$.

Insertion of Eq. (13) into Eq. (3) yields the following straightforward expression for $\chi_o$, as a function of chain length

$$\chi_o = \alpha - \zeta\left(\frac{1}{2} + \kappa N^{-(1-a)}\right) \quad (14)$$
$$\text{with} \quad \alpha = \chi_{o,f-c}$$

In order to transform this relation to describe the situation for the directly accessible second osmotic virial coefficient, we insert Eq. (14) into Eq. (2), resolve the resulting expression with respect to $A_2$ and separate $N$-dependent from $N$-independent members. The obtained relation reads

$$A_2 = A_2^\infty + \sigma N^{-(1-a)} \quad (15)$$

where

$$A_2^\infty = \frac{1 + \zeta - 2\alpha}{2\rho_2^2 \overline{V_1}} \quad (16)$$

and

$$\sigma = \frac{\zeta \kappa}{\rho_2^2 \overline{V_1}} \quad (17)$$

The general suitability of Eq. (15) for the representation of actual data concerning the chain length dependence of $A_2$ has already been verified[5]. As postulated by Eq. (16) the second osmotic virial coefficients does not generally become zero in the limit of infinitely long chains but only for $\zeta = 2\alpha - 1$. For theta solvents this condition is always fulfilled because $\alpha = 0.5$ and $\zeta = 0$. $A_2^\infty$ can be positive or negative, its sign depends on the relative magnitudes of $\alpha$ and $\zeta$. Whether an increase in the chain length of the polymer results in a diminution (as usual) or in a augmentation of $A_2$ depends according the present approach on this sign of $\zeta$, the conformational response of the polymer backbone to dilution.

The relations presented above provide an opportunity to calculate $\zeta$ and $\alpha$ from the chain length dependence of $A_2$. Rearrangement of Eq. (17) yields

$$\zeta = \frac{\sigma \rho_2^2 \overline{V_1}}{\kappa} \quad (18)$$

and Eq. (16) gives

$$\alpha = \frac{1 + \zeta}{2} + A_2^\infty \rho_2^2 \overline{V_1} \quad (19)$$

## B. Quantitative considerations

In the following we evaluate the experimental information on $A_2$ ($N$) according to the present approach. Due to the vastly different solvent qualities of MEK, TL and n-Oct for PDMS, the exponents $a$ of the Kuhn-Mark-Houwink relation span the range from 0.5 to 0.8, and consequently the independent variable $N^{-(1-a)}$ extends over very different intervals for the given set of samples, unlike the conventional evaluation of $A_2$ ($M$) data (cf. Figure 3). For this reason we present two graphs. The upper part of Figure 6 contains the data for the marginal solvent MEK and for TL, representing a typical good solvents, whereas the results for the extremely good solvent n-Oct are shown in the lower part of Figure 6.

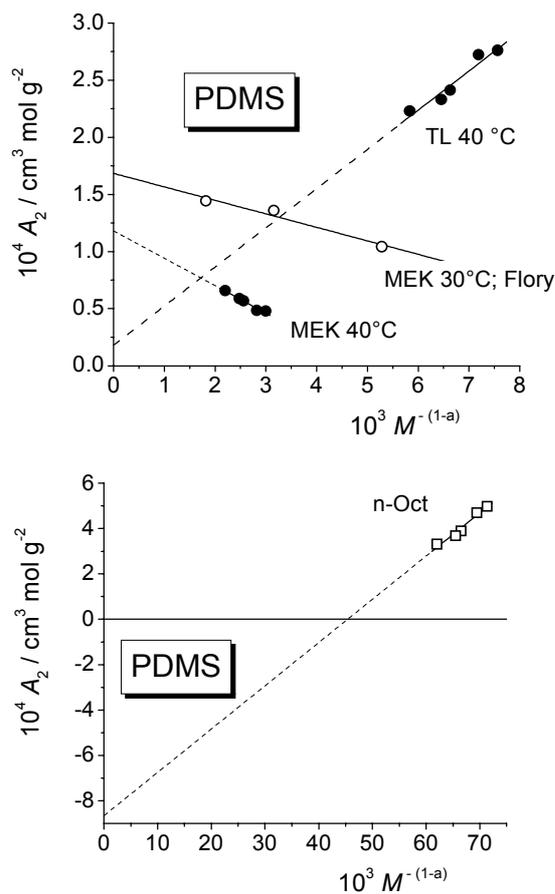

Figure 6. Representation of $A_2$ as a function of $M$ according to Eq. (15). *Upper part*: MEK/PDMS and TL/PDMS; *lower part*: n-Oct/PDMS.

Like all $A_2$ ($N$) data evaluated according to Eq. (15) so far[5], the present results yield straight lines, where the intercept is in the general case different from zero. At 40 °C, somewhat above the theta-temperature (20 °C) of the system MEK/PDMS, $A_2$ rises as $M$ becomes larger, i.e. $\chi_0$ decreases. This finding is a further example for the principle inadequacy of theoretical concepts postulating that the second osmotic virial coefficient should inevitably become smaller with increasing chain length of the polymer[3]. It is worthwhile to note that the evaluation of the early osmotic pressure data reported[4] by Flory et al. with respect to $A_2$ and their analysis according to Eq. (15) corroborates the present findings, as demonstrated in Figure 6. The qualitative agreement of the two sets of data is beyond doubt. However, the fact that Flory's $A_2$ values are larger at 30 °C (i.e. closer to $\Theta$) than ours at 40 °C requires discussion. A rationalization via a contamination of the solvent in one of the experiments can be excluded in view of the consistent Kuhn-Mark-Houwink relations (Figure 5). Most likely the explication lies in the dissimilar molecular non-uniformities of the polymer samples in conjunction with the fact of different averages of $A_2$ and $M$ in osmotic and in light scattering experiments. This tentative explanation of the discrepancies is backed by the joint evaluation of the different results in Figure 7, where all data fall on a common line.

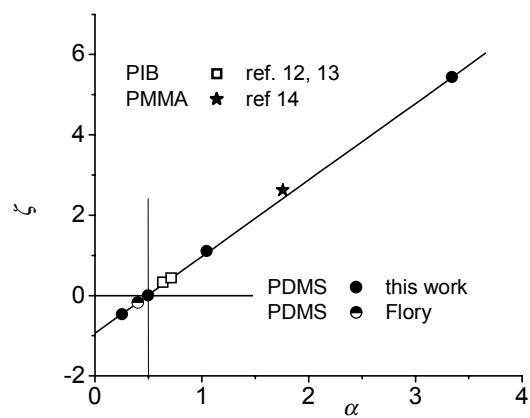

Figure 7. Conformational response, $\zeta$, as a function of $\alpha$, the interaction parameter for fixed conforma-

tion. In addition to the present results this graph also shows data[4] for MEK/PDMS at 30 °C and for some solutions of polyisobutylene[12,13].

Slopes and intercepts of the lines of Figure 6 yield $\zeta$ (the conformational response) and $\alpha$ (the interaction for fixed conformation) by means of the Eqs. (18) and (19). The resulting data are collected in Table 2. They demonstrate, like all previous findings[5], that solvents are thermodynamically favorable because of the conformational relaxation $\zeta$, despite unfavorable, i.e. positive, $\alpha$ values. This observation appears surprising at first. However, on second thought it is not unreasonable if one keeps in mind that the Gibbs energy of dilution at fixed conformation (dominated by enthalpy) should remain low because the effect must be distributed on the many segments involved in the opening of an intermolecular contact. The main reduction of the Gibbs energy of the system is associated with the conformational rearrangements (dominated by entropy).

**Table 2**. Collection of the parameters required for the evaluation of the molecular weight dependencies of the second osmotic virial coefficients of PDMS in the indicated solvents.

|  | MEK 20 °C | MEK 40 °C | TL 40 °C | n-Oct 40 °C |
|---|---|---|---|---|
| $10^3 K_w$ / mL g$^{-1}$ | 63.20 | 50.80 | 31.70 | 3.51 |
| $a$ | 0.50 | 0.56 | 0.63 | 0.80 |
| $\kappa$ / mL g$^{-1}$, Eq. (13) | 0.563 | 0.597 | 0.561 | 0.194 |
| $10^4 \sigma$ / cm$^3$ mol g$^{-2}$, Eq. (15) | 0.00 | - 33.35 | 62.25 | 69.15 |
| $10^4 A_2^\infty$ / cm$^3$ mol g$^{-2}$, Eq. (15) | 0.00 | 1.18 | 0.19 | - 8.65 |
| $\alpha$, Eq. (14) | 0.500 | 0.254 | 1.050 | 3.347 |
| $\zeta$, Eq. (14) | 0.000 | - 0.471 | 1.104 | 5.430 |

From general experience it is known[2] that the absolute values of the enthalpy contributions and that of the entropy contributions to the Gibbs energy are normally one order of magnitude larger than the absolute value of the Gibbs energy itself. Furthermore, all favorable changes in entropy (or enthalpy) are normally accompanied by a corresponding unfavorable change in enthalpy (or entropy). With regard to the interpretation of $\alpha$ and $\zeta$ in terms of enthalpy and entropy, these two parameters ought to be considerably larger than $\chi_o$ and interrelated. Figure 7 demonstrates that this is indeed the case. Despite the discrepancy in the absolute values of $A_2$ reported by Flory and coworkers for MEK at 30 °C and our values for 40 °C the data fall on the same line. The conformational response $\zeta$ passes zero at $\alpha$ = 0.5 as the theta-conditions are reached and $A_2$ becomes zero, independent of chain length. In order to gain information on the behavior of further polymers we have incorporated the results of an earlier evaluation[5] of data published for solutions of polyisobutylene (PIB) in two different solvents[12,13]. In addition we have evaluated data for solutions of poly(methyl methacrylate) (PMMA) in acetone[14]. It is interesting to note that these points are very close to the relation for PDMS; this is probably so because of a rather similar conformational behavior of the two types of macromolecules. For stiffer polymers with lower conformational variability the room for conformational response to dilution diminishes and the contributions of $\zeta$ to $\chi_o$ should become less and the line shown in Figure 7 needs no longer hold true.

The observation of negative $\zeta$ values might surprise at first view because it seemingly contradicts the idea of a spontaneous conformational relaxation. However, it is not the chemical potential of the solvent (the basis of $\zeta$) but the Gibbs energy of the entire system that must become minimum under equilibrium conditions. Up to now we have not explicitly addressed the reasons for this atypical negative $\zeta$ values. In case of the solutions of a liquid crystalline polymer[15] the explanation could well lie in a rather low chain flexibility, hampering conformational relaxation, as discussed above. The other examples refer to solutions of polystyrene (PS) in tertiary butyl acetate (TBA) close to the *exothermal* theta temperature of the system[16] and to solutions of PDMS in MEK close to the *endothermal* theta temperature. In both cases the distance to $\Theta$ is low and one might therefore speculate that the vicinity to theta conditions is essential. This option should, however, be ruled out in view of the fact that reports on the above behavior are scarce, despite abundant measurements under near theta conditions. The most probable explanation lies in particular features of the polymer/solvent system. With the TBA/PS the peculiarity lies in the highly expanded structure of the solvent at the elevated temperatures associated with the exothermal theta conditions. Under these circumstances the conformational response to solvent addition becomes negative because of the tendency of the segments to search for another segment after dilution in order to keep away from the numerous voids representing in their totality the large free volume of the system. With MEK/PDMS the situation is similar, but this time because of the very favorable intersegmental contacts PDMS can form due to its very flexible molecular architecture[17,18]. The obvious correlation of conformational relaxation with the composition dependence of chain dimensions leads to the supposition that the coils should shrink upon dilution, despite the fact that these solvents are still good enough to guarantee complete miscibility of the components. Another finding that can be tentatively rationalized in molecular terms are the uncommonly large $A_2$ values for the system n-Oct/PDMS. According to the current approach the extremely high solvent quality results from a very favorable conformational relaxation. It is not unreasonable to assume that it is due to the linear molecular architecture of the solvent. The addition of n-Oct to the solution at fixed chain conformation ought to lead to a rather limited number of favorable contacts between its $-CH_2-$ or $CH_3-$ groups and the now separated polymer segments. Conformational relaxation may rise the number of these advantageously interacting sites markedly.

## V. CONCLUSIONS

This contribution demonstrates that the explicit provision for two typical features of linear flexible macromolecules - chain connectivity and conformational variability - leads to a simple expression for the chain length dependence of the Flory-Huggins interaction parameter in the limit of infinite dilution. This relation describes all experimental findings concerning $A_2$ ($N$) quantitatively in a clear-cut manner with a minimum of physically meaningful parameters. Above all it enables the rationalization of the existence of systems for which $A_2$ *increases* (i.e. $\chi_o$ *decreases*) with rising chain length, in contrast to the expectation of all current theories. Furthermore the approach accounts for non-zero $A_2$ values ($\chi_o \neq 0.5$) in this limit.

Due to the limited experimental material it was so far impossible to deal with some interesting problems that came into view during this investigation. In particular it would be informative to break up the parameters $\alpha$ and $\zeta$ into their enthalpy and entropy contributions, by studying temperature influences, in order to learn more on the ways the compromise between enthalpic gains and entropic losses is achieved. Another aspect that deserves further consideration concerns the effect of chain flexibility on the extent of conformational response, where the limit of very stiff chains appears particularly interesting.

Even more rewarding than the answers to the above questions is an extension of the present approach into the region of high polymer concentrations. The necessity of an improvement of our understanding regarding the composition dependence of the Flory-Huggins interaction parameter is clearly demonstrated by numerous experimental findings that cannot be rationalized by current theories, like the existence of pronounced minima[2] in $\chi$ ($\varphi$). Part II of the publication is therefore dedicated to the generalization of the present approach to cover the entire composition range. In contrast to all previous attempts to describe the composition dependence of $\chi$ it does not start from the concentration regime of large coil overlap but from dilute solutions within the composition range of pair interaction between only two individual macromolecules.

## ACKNOWLEDGEMENTS

We are grateful to the Deutsche Forschungsgemeinschaft for financial aid. Furthermore we would like to thank the Romanian Academy of Science, which supported us within the frame of an interacademic project.

**References**

[1] P. J. Flory, *Principles of Polymer Chemistry* (Cornell University Press, Ithaca USA, 1953).
[2] H.-M. Petri, N. Schuld, and B. A. Wolf, Macromolecules **28**, 4975 (1995).
[3] H. Yamakawa, Polym. J. **31**, 109 (1999) and literature cited therein.
[4] P. J. Flory, L. Mandelkern, J. B. Kinsinger, and W. B. Shultz, J. Am. Chem. Soc. **74**, 3346 (1952).
[5] B. A. Wolf, Makromol. Chem. **194**, 1491 (1993).
[6] S. W. Kantor, W. T. Grubb, and R. C. Osthoff, Journal of the Amercian Chemical Society **76**, 5190 (1954).
[7] P. R. Dvornic, R. W. Lenz in High Temperature Siloxane Elastomers, *Chap. II- Polysiloxanes* (Hüthig & Wepf Verlag Basel, Heidelberg, New York, 1950) 25.
[8] J. Chojnowski, *J. Inorg. Organomet. Polym.* **1**, 299 (1991).
[9] A. Hinrichs and B. A. Wolf, Macromol. Chem. Phys. **200**, 368 (1999).
[10] F. Müller-Plathe, H. Liu, and W. F. van Gunsteren, Computational Polymer Science **5**, 89 (1995).
[11] B. A. Wolf, Adv. Polym. Sci. **10**, 109 (1972).
[12] H. Geerissen, P. Schützeichel, and B. A. Wolf, Macromolecules **24**, 304 (1991).
[13] W. R. Krigbaum, Journal of the Amercian Chemical Society **76**, 3758 (1954).
[14] R. Kirste and G. Wild, Makromol. Chem. **121**, 174 (1969).
[15] H. Krömer, R. Kuhn, H. Pielartzik, W. Siebke, V. Eckhardt, and M. Schmidt, Macromolecules **24**, 1950 (1991).
[16] B. A. Wolf and H. J. Adam, J. Chem. Phys. **75**, 4121 (1981).
[17] P. J. Flory, *Statistical mechanics of chain molecules* (John Wiley & Sons, New York, USA, 1969).
[18] A. Muramoto, Polymer **23**, 1311 (1982).